\documentclass[a4paper, oneside, twocolumn, notitlepage, 10pt]{extarticle_ecoc}
\usepackage{ecoc}
\usepackage{comment}

\begin{document}
\selectlanguage{english}    


\title{Real-Time Monitoring of Cable Break in a Live Fiber Network using a Coherent Transceiver Prototype}%


\vspace{-0.8cm}
\author{
    Mikael~Mazur\textsuperscript{(1)}, 
    Lauren~Dallachiesa\textsuperscript{(1)}, 
    Roland~Ryf\textsuperscript{(1)}, 
    Dennis~Wallberg\textsuperscript{(2)}, 
    Erik~B{\"o}rjeson\textsuperscript{(3)}, \\ 
    Magnus~Bergroth\textsuperscript{(2)}, 
    B{\"o}rje Josefsson \textsuperscript{(2)} 
    Nicolas~K.~Fontaine\textsuperscript{(1)}, 
    Haoshuo~Chen\textsuperscript{(1)}, 
    David~T.~Neilson\textsuperscript{(1)}, \\
    Jochen~Schr{\"o}der\textsuperscript{(4)}, 
    Per~Larsson-Edefors\textsuperscript{(3)} and 
    Magnus~Karlsson\textsuperscript{(4)} \vspace{-3mm} 
}
\maketitle                  
\begin{strip}
 \begin{author_descr}
 
\textsuperscript{(1)} Nokia Bell Labs, 600 Mountain Ave., Murray Hill, NJ 07974, USA \\
\textsuperscript{(2)} Sunet, Tulegatan 11, Stockholm, Sweden \\
\textsuperscript{(3)} Dept. of Computer Science and Engineering, Chalmers University of Technology, Sweden \\
\textsuperscript{(4)} Dept. of Microtechnology and Nanoscience, Chalmers University of Technology, Sweden
 \end{author_descr}
 \vspace{-2mm}
\end{strip}
\setstretch{1.1}
\renewcommand\footnotemark{}
\renewcommand\footnoterule{}


\begin{strip}
  \begin{ecoc_abstract}
We monitor a 524\,km live network link using an FPGA-based sensing-capable coherent transceiver prototype during a human-caused cable break.  Post-analysis of polarization data reveals minute-level potential warning precursors and baseline-exceeding changes directly preceding the break.
~\textcopyright2023 The Author(s)\end{ecoc_abstract}
\vspace{-3mm}
\end{strip}

\section{Introduction}
\label{sec:intro}
Lately, there has been a large interest in sensing use active telecom, both for environmental monitoring and to improve the network performance\cite{ip2022distributed}. 
Fiber sensing using deployed telecom fibers is typically performed using tailored sensing methods such as distributed acoustic sensing (DAS)\cite{Daley2013} or laser phase interferometry\cite{Marra2018}. 
Examples of sensing using DAS and submarine cables include close to shore ocean monitoring and earthquake detection~\cite{Lindsey2019} and whale migration ~\cite{landro2022sensing}. 
Demonstrations using terrestrial networks include monitoring remote seismic activity~\cite{Martins2019} and traffic\cite{Wellbrock2021}. However, while these techniques provide excellent sensitivity dedicated fibers/wavelengths are required. 

\begin{figure*}[ht]
   \centering
    \includegraphics[width=1\linewidth]{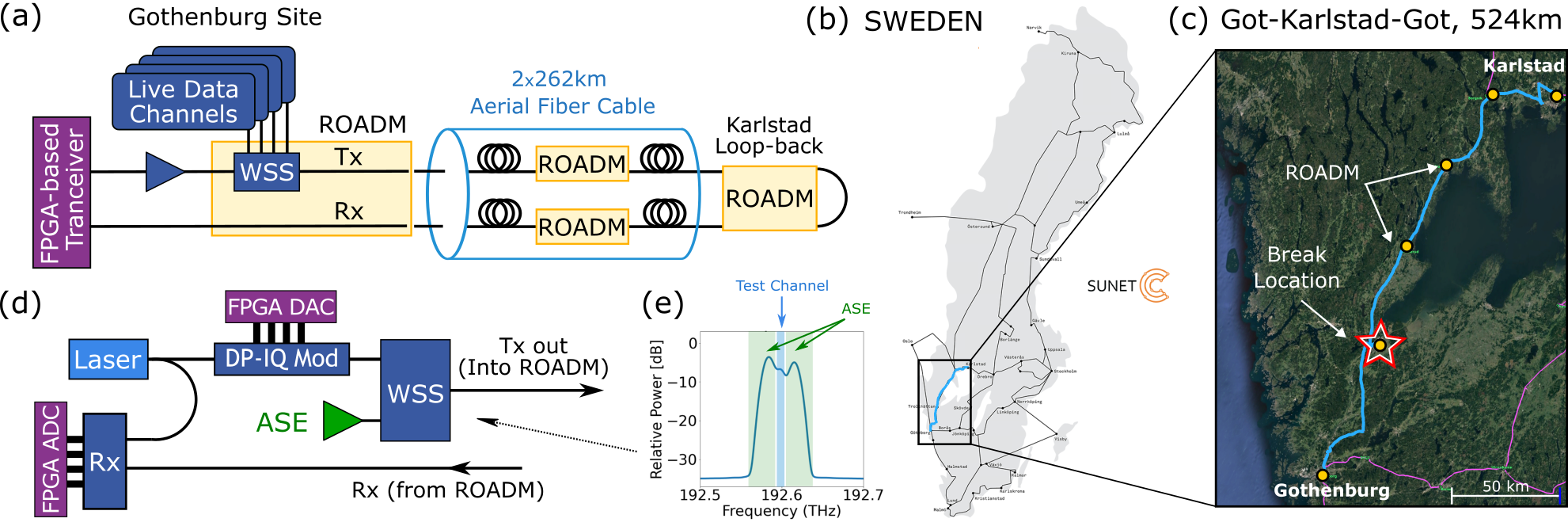}
    \caption{(a) Experimental setup illustrating the network layout and how the FPGA-based coherent transceiver prototype was connected to the network for continuous monitoring/sensing. 
    (b) A map of Sunet's Swedish fiber network with the 524\,km route round-trip from Gothenburg to Karlstad and back highlighted in blue. (c) Zoom-in of the route consisting of aerial fiber, including five ROADM nodes, which were connected with the aerial cables using shorter segments of buried fiber. The fiber break happened as one of these segments was accidentally exposed by an excavator, with the approximate location illustrated by the star. (d) Detailed sketch of the FPGA-based coherent transceiver prototype with enhanced sensing capabilities. (e) A 50\,GHz channel was emulated by combining the FPGA-based transceiver wavelength with ASE in a WSS with 6.25\,GHz address-ability and 1 wavelength slot devoted to the test signal and the remaining 43.75\,GHz filled with ASE.}
    \label{fig:setup}
\end{figure*}

As a complement to dedicated fiber sensing methods, the use of coherent transceivers has been proposed\cite{Zhan2021,Mazur2022}.
Sensing using coherent transceivers has the inherent advantage of full compatibility with telecom systems and avoids any loss of bandwidth. 
With transceivers, the ``sensor" integrates the network path and is subject to optical routing. 
It also implies that a wavelength channel is inherently covered by network sensing, independent of if dedicated per-span fiber sensors are present or not. 
Monitoring can be done by following the state-of-polarization (SOP), which is continuously tracked by the digital signal processing (DSP) in the receiver\cite{Faruk2017,Crivelli2012}. This has been demonstrated using coherent transceivers for submarine~\cite{Zhan2021}, terrestrial~\cite{Mazur2022b} and aerial fiber systems~\cite{Mazur2023}. Polarization sensing is also independent of laser linewidth~\cite{Mecozzi2021}. Transceiver-based phase sensing\cite{ip2022distributed,Mazur2022} and DSP-based time-of-flight measurements~\cite{Mazur2023} have also been demonstrated. 
Polarization-based sensing directly targeting proactive fiber break detection has been investigate inside the lab over short fibers subjected to robotic movements using both coherent transceivers~\cite{Boitier2017} and optical supervisory channels~\cite{Simsarian2017}. 
If a network sensor, or active transceiver, detects baseline deviations prior to the actual break, it could trigger an alarm and with sufficiently long warning, breaks might be avoided. 
However, to the best of our knowledge, no transceiver-based measurements of an active cable break, including the presence of potential pre-cursors, have been reported. 

In this work, post-factum analysis of results captured using coherent receiver monitoring in a live network during a fiber break event. The break was caused by an excavator accidentally exposing the fiber cable during construction work. The 524\,km link includes 5 ROADMs and consists mainly of aerial fiber, with shorter pieces of buried fiber connecting the ROADM nodes to the power line locations. 
The fiber was monitored using an FPGA-based coherent transceiver prototype co-propagated along the live coherent data channels present on the network. 
We show polarization-based sensing of baseline data for the link together with the changes occurring during the actual break. Our analysis shows that for this specific fiber break, the associated polarization stands out from the baseline with noticeable features reaching 50\,Hz. 
A few potential precursors are observed minutes before the actual break. Despite the uncontrolled nature of live fiber experiments, these polarization changes are significantly stronger than the baseline observed over weeks. 
Our results show that there is potential for fiber transceiver-based sensing to take an active role in improving the stability of future fiber networks.

\section{Experimental Setup}\vspace{-1mm}
The experimental setup is shown in Fig.~\ref{fig:setup}(a). The FPGA-based prototype transceiver was connected to Sunet's (Swedish University Network's) live fiber network at the reconfigurable add-drop multiplexer (ROADM) node in Gothenburg, Sweden.  Multiple live coherent transceivers were also connected through the same wavelength selective switch (WSS) input. A map of Sunet's network is shown in Fig.~\ref{fig:setup}(b) with a zoom-in of the route used shown in Fig.~\ref{fig:setup}(c). The route is highlighted in blue and consisted exclusively of aerial fiber. 
The aerial fiber is only temporarily routed to ground in order to pass through ROADM nodes. Throughout the link, 5 ROADM nodes were passed before reaching Karlstad. The fiber break happened as the buried cable  was accidentally exposed by an excavator during construction in the proximity of one of the ROADM nodes. The approximate break location is illustrated by the star in Fig.~\ref{fig:setup}(c). 
A detailed sketch of the transceiver prototype is shown in Fig.~\ref{fig:setup}(d). A single FPGA with integrated DACs and ADCs (Xilinx ZU48DR) was used to implement the real-time transceiver module. The digital transmitter operated with a parallelism of 16 at a clock rate of 125\,MHz to generate a two-fold over-sampled 1-GBd signal. 
A 50\,GHz wide channel was emulated by combining the narrow-band prototype signal with amplified spontaneous emission (ASE) noise in a commercial WSS with 6.25-GHz resolution, as shown in Fig.~\ref{fig:setup}(e). 
The loop-back in Karlstad was implemented using the ROADM node, routing our channel back to Gothenburg where the drop node of the ROADM was used to extract the test signal.
Full polarization sensing was enabled by extracting the internal equalizer state up to MHz rate. 
To enable long-term measurements without exhausting storage capacity, a sampling period of \(\sim \)100\,\textmu s was selected. A 3-dimensional Stokes vector was calculated from the received Jones matrices. The experimental setup is further described in~\cite{Mazur2023}. 

\begin{figure*}[t]
   \centering
    \includegraphics[width=1.\linewidth]{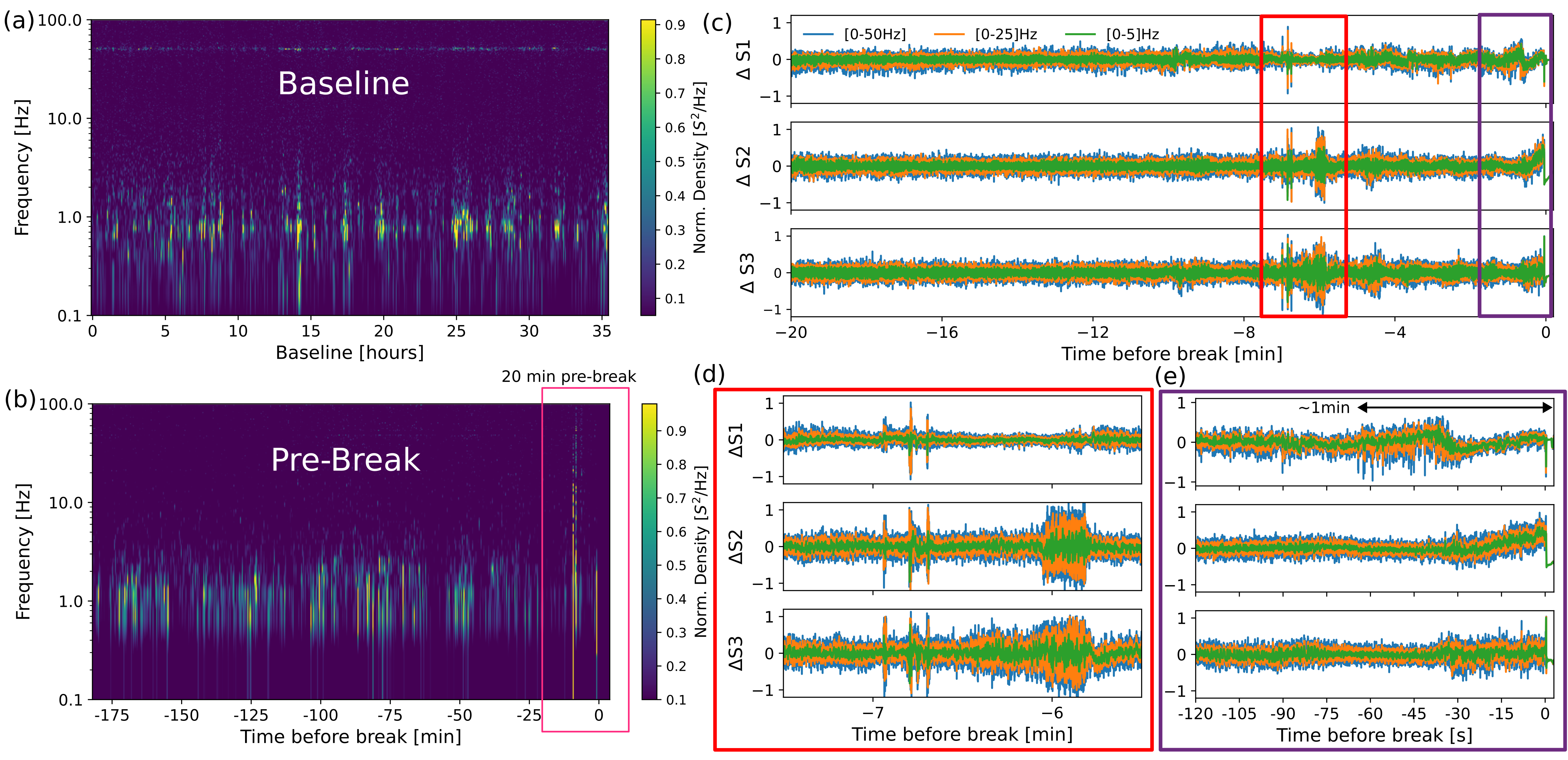}
    \caption{(a) Baseline spectrogram for S2 during a 36h example window two days preceding the break. Long-term analysis showed similar baselines over weeks, with larger frequency changes around sub-Hz to Hz level, matching well with environmental changes affecting the long aerial fiber link. The aerial link is highly susceptible to both wind and temperature changes. (b) Spectrogram for three hours preceding the break with the 50\,Hz contribution filtered out. We observe an increased intensity in the range of 0.1-1\,Hz. Importantly, about 10 minutes prior to the break, a large impulse-like change incorporating frequencies up to, and above, 50\,Hz can be observed. (c) Corresponding deviations for S1/S2/S3 for the last 20 minutes preceding the break. This corresponds to rotational changes from the initial state-of-polarization. (d) Zoom-in of the potential strong precursors occurring about 6 minutes prior to the break.  (e) Zoom-in of the actual break with distinct changes in the polarization behavior observed around 1 minute prior.}
    \label{fig:res2}
\end{figure*}

\vspace{-3mm}
\section{Results}
The monitored SOP shows strong presence of 50\,Hz and harmonics, which were filtered out for the event analysis. These oscillations likely originates from the Faraday effect arising from the aerial part of the fiber was spun around high voltage conductors in the backbone power grid~\cite{Mazur2023}. The high SOP sampling rate of about 10kHz enabled this without suffering from aliasing effects. 
A spectrogram for the filtered Stokes parameter S2 during a 36-hour example baseline, starting two days prior to the break, is shown in Fig.~\ref{fig:res2}(a). We observe that most of the energy is located within $<$50~Hz window, matching well with previous results. The impact of the 50\,Hz tone is also visible. The weak, slowly-varying signal behavior is also seen in Fig.~\ref{fig:res2}(a) showing the steady-state baseline observed over longer time scales.  The observed rate-of-change also matches well with environmental effects, such as wind, which is affecting the aerial fiber~\cite{Mazur2023}. 
Figure~\ref{fig:res2}(b) shows the corresponding spectrogram for the last three hours prior to the cable break. The 50\,Hz tone has been filtered out with a 1.5\,Hz wide bandpass filter. Here we observe a somewhat increased background level, but no major high-frequency components are presents until the very last moment before the break. 
Figure~\ref{fig:res2}(c) shows the spectrally filtered changes for S1/S2/S3 during the 20-minute time period leading up to the fiber break. A few interesting observations can be made. First, around 5-7 minutes before the break, we observe stronger changes in the SOP. The exact nature of these changes is unknown, but it is likely they are related to the ground construction activities that in the end lead to the fiber break. Worth noticing here is that these fluctuations are stronger than any other fluctuations observed during the week-long window before the actual break. 
Secondly, these events have noticeably higher frequency content compared to the baseline, which is clearly visible on the spectrogram in Fig.~\ref{fig:res2}(a). 
A zoom-in of the events around 7~minutes before the break is shown in Fig.~\ref{fig:res2}(d). Here we see that the first set of events seems to be higher intensity and more transient in nature, giving sharper responses separated by around 10~s. For the events around 6~minutes we observe an event that is different in nature, and seems to give more rapid fluctuations for about 15~s. The event is most evident for S2. Given the uncontrolled nature of this live fiber experiment, tracing back the exact origin is not feasible and details of the construction procedure leading up to the break are not available. Still, the results in Fig.~\ref{fig:res2}(d) show that SOP monitoring has a potential to identify events that are disruptive in nature, some of which could potentially provide an early-warning alarm. 
A zoom-in of the break interval is shown in Fig.~\ref{fig:res2}(e). Here we see slower, but significant changes leading up to the break. The break itself shows up as spikes prior to loss of signal with frequency content exceeding 100\,Hz at ms time scale. 

\section{Conclusion}
We have presented polarization sensing data collected from a real-time FPGA-based coherent transceiver prototype during a live network fiber break. We compared long-term baseline data to the events leading up to the break, highlighting potential precursors and stronger changes in the state-of-polarization occurring on a timescale of minutes to seconds prior to the actual cable break. While the baseline measurements show that most polarization changes occur at frequencies around 1\,Hz, matching well with environmental changes, the potential precursors stand out and contain higher-frequency content reaching 50\,Hz. Our results show the potential for coherent transceivers with enhanced monitoring capabilities to play a vital role in improving the stability of future optical networks by enabling proactive rerouting and potential break avoidance using early-warning signatures extracted from the coherent digital signal processing.




\clearpage

\printbibliography
\end{document}